\documentclass[%
 reprint,
 amsmath,amssymb,
 10pt,english, aps,superscriptaddress,showpacs,floatfix,prl,lengthcheck,floatfix,nofootinbib,twocolumn
]{revtex4-1}

\usepackage{graphicx}
\usepackage{dcolumn}
\usepackage{bm}
\usepackage[pdfusetitle,colorlinks,linkcolor=blue,citecolor=blue,urlcolor=blue]{hyperref}

\begin{document}

\preprint{APS/123-QED}

\title{Reconstruction of attosecond beating by interference of two-photon interband transitions in solids}

\author{Rui E. F. Silva}
 \affiliation{Instituto de Ciencia de Materiales de Madrid (ICMM), Consejo Superior de Investigaciones Cient\'ificas (CSIC), Sor Juana In\'es de la Cruz 3, 28049 Madrid, Spain}
\author{\'Alvaro Jim\'enez-Gal\'an}%
 \email{jimenez@mbi-berlin.de}
\affiliation{%
 Max-Born-Institute, Max-Born Stra{\ss}e 2A, D-12489 Berlin, Germany.}%
\affiliation{
 Joint Attosecond Science Laboratory, National Research Council of Canada and University of Ottawa, Ottawa, Canada.}%

\begin{abstract}
The reconstruction of attosecond beating by interference of two-photon transitions (RABBIT) is one of the most widely used techniques for obtaining both the relative phases of harmonics forming an attosecond pulse train and the phase of atomic radiative transitions. If the latter is computed by theory, it allows to reconstruct the attosecond pulse train; if the former is known experimentally, it allows reconstruction of the electronic dynamics of photoionization in atomic and molecular systems with attosecond temporal resolution. As it relies on the interference of photo-electrons in vacuum, similar interference has never been contemplated inside crystals. Here we explore the applicability of this scheme to solid-state systems using a one-dimensional model and a DFT-calculated structure of 2D hexagonal boron nitride. We discuss the possibility of: (i) reconstructing the relative phases between harmonics with trivial influence of the ``atomic phase”, (ii) retrieving the relative phases of two-photon transitions through different bands, which are generally challenging to obtain both experimentally and numerically. These phases are recorded in the beating of the population signal arising from interfering two-photon pathways, and can be read-out with angle-resolved photo-emission spectroscopy. Furthermore, the amplitude of the population beating decays as the pump and probe pulses are separated in time due to electron-hole decoherence, providing a simple interferometric method to extract dephasing times.
\end{abstract}

\maketitle


\section{Introduction}

Advances in ultrafast laser technology during the last two decades have given rise to the field of attosecond science - the study and control of electron dynamics at their natural (attosecond) timescale~\cite{Goulielmakis2007, Krausz2009}. Among the experimental techniques that made this possible, the reconstruction of attosecond beating by interference of two photon transitions (RABBIT) stands out~\cite{Veniard1996, Muller2002, Paul2001, Klunder2011, Gruson2016, Donsa2019}. Since it uses weak electric fields, it allows to monitor coherent electronic dynamics that are barely modified by the laser. RABBIT relies on the controlled interference of quantum paths in a pump-probe scheme. It can be implemented with any combination of four frequencies provided the two-photon paths created by the sequential absorption/stimulated emission of the frequencies reach the same final energy.

In its usual implementation, the quantum paths are created by two harmonics $(2N\pm 1)\omega$ of a frequency comb (the pump) that photo-ionize an initial bound electronic state (Fig.~\ref{fig:hBN}a). The harmonic comb is generated using a strong fundamental field of frequency $\omega$ on an inversion symmetric target, thus guaranteeing that no even ($2N\omega$) harmonics are present. The interference of the paths containing the $(2N\pm 1)\omega$ photons is controlled by time-delaying a weak, phase-locked replica of the generating $\omega$ field (the probe) that creates sidebands at $\Omega = 2N\omega$ energies whose intensity oscillates as a function of the pump-probe time delay $\tau$~\cite{Veniard1996, Paul2001, Jimenez2016},
\begin{equation}\label{eq:beating}
I_{\text{SB}} (2N\omega) \propto \cos (2\omega \tau + \theta_{2N}).
\end{equation}
The equation above is the characteristic equation of RABBIT, computed using second-order perturbation theory (see Appendix C). We use atomic unit throughout unless otherwise stated. Eq.~\ref{eq:beating} relates the measured observable, i.e., the sideband beating phase $\theta_{2N} = \Delta \phi_{2N} + \Delta \varphi_{2N}$, to the phase difference between adjacent harmonics in the comb, $\Delta \phi_{2N} = \phi_{2N-1} - \phi_{2N+1}$, and the so-called atomic phase $\Delta \varphi_{2N}$. For long pulses, the latter corresponds to the relative phase of the two-photon matrix elements~\cite{Dahlstrom2012, Jimenez2016},
\begin{equation}\label{eq:atomic_phase}
\begin{split}
\Delta \varphi_{2N} &= \varphi_{2N+1} - \varphi_{2N-1} = \\
& \arg \left\{\mathcal{M}_{nm}\left((2N+1)\omega\right) + \mathcal{M}_{nm} \left(-\omega\right) \right\} - \\
&\arg \left\{\mathcal{M}_{nm}\left((2N-1)\omega\right) + \mathcal{M}_{nm} \left(\omega\right) \right\},
\end{split}
\end{equation}
where $\mathcal{M}_{nm} (\omega) = \sum_{j}\mathcal{O}_{nj}\mathcal{O}_{jm} / (E_m + \omega - E_j)$ is the two-photon matrix element, $\mathcal{O}_{nm}$ is the dipole operator, e.g., $\mathcal{O}_{nm} = \langle n | \mathbf{r} | m \rangle$ in the length gauge, and $E_m$ is the energy of the initial state. Initially, RABBIT was applied to characterize the relative phases of harmonics in attosecond pulse trains by assuming that the atomic phase is a smooth function of energy~\cite{Paul2001}. Later, its application shifted towards the reconstruction of the relative amplitudes and phases of photoionization matrix elements from different orbitals, including transitions through autoionizing states~\cite{Klunder2011, Dahlstrom2012, Pazourek2015, Isinger2017, Argenti2017, Vos2018}. This has allowed to extract photo-ionization time delays and reconstruct the temporal evolution of correlated electronic wavepackets~\cite{Gruson2016, Busto2018} - two hallmarks of attosecond science.

In the last decade, attosecond science has advanced from atomic and molecular targets towards condensed matter systems~\cite{Kruchinin2018}. Techniques such as high harmonic spectroscopy~\cite{Ghimire2011}, high sideband generation~\cite{Langer2018}, attosecond streaking~\cite{Cavalieri2007, Garg2016} and x-ray absorption spectroscopy~\cite{Moulet2017,Buades2018} have been implemented in solids, allowing e.g., to image valence potentials with picometer resolution~\cite{Lakhotia2020}, track metal-to-insulator~\cite{Silva2018} and topological phase transitions~\cite{Bauer2018, Silva2019}, characterize inelastic scattering time in dielectrics~\cite{Seiffert2017}, or control and measure the valley degree of freedom~\cite{Langer2018, Jimenez2019}. The RABBIT technique has been used to perform time-resolved photoemission experiments from solid surfaces, i.e., surface RABBIT~\cite{Locher2015, Ambrosio2018}, where the time delay of electrons emitted from noble gas surfaces to the photoionization continuum was extracted.

The RABBIT technique relies on the interference of the two-photon paths in the photo-electron continuum and therefore has not yet been applied to study the dynamics inside solids, i.e., transitions between bands. Here, we show that the electron momentum distribution of a band populated through two-photon resonant interband transitions displays the characteristic RABBIT beating and allows to record information on the relative phase of the harmonics forming the attosecond pulse train and the Berry connections between the bands. This observable can be extracted with standard angle-resolved photo-emission spectroscopy (ARPES). Since ARPES is a one-photon process, the relative phase of the interband transitions, i.e., the RABBIT beating signal, remains unaffected by the measurement. Akin to transitions between autoionizing resonances in atoms~\cite{Jimenez2014,Jimenez2016,Gruson2016}, the RABBIT beating in interband transitions remains even when the pump and probe pulses do not overlap in time. As a function of the pump-probe time delay, the beating amplitude decays as a consequence of electronic decoherence, and can thus be used to measure dephasing times.

\section{Results}

In this work we will concentrate on extracting dynamical information of transitions between bands that are close to the gap of semiconductor materials. The energy separation between these bands is usually on the order of few electronvolts, so that it is in principle possible to use the scheme we propose below using typical laser wavelengths.

\subsection*{Reconstruction of the relative harmonic phase}

\begin{figure*}
\begin{center}
\includegraphics[width=\linewidth]{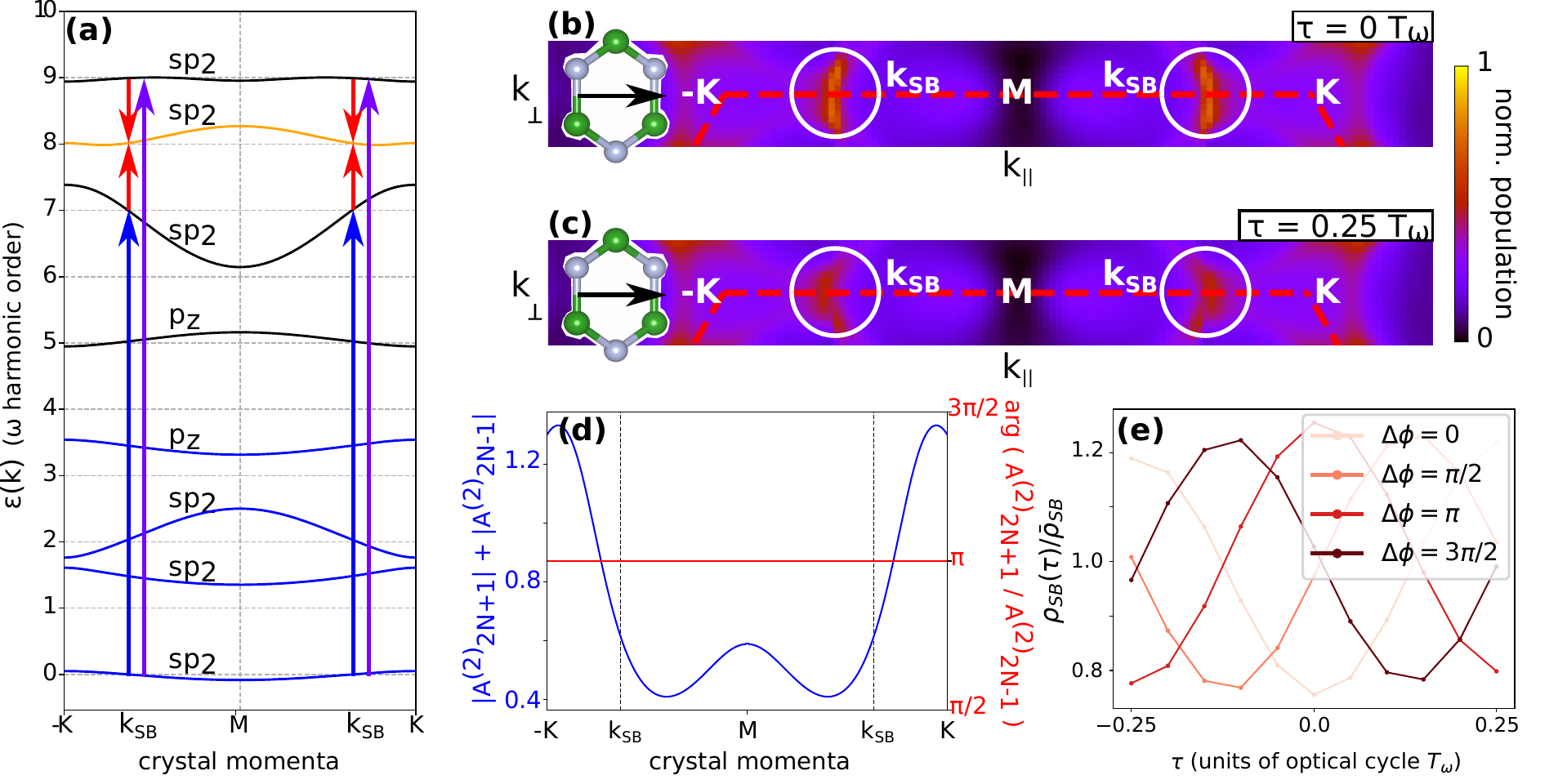}
\caption{\label{fig:hBN} (a) Energy dispersion of hBN. The vertical axis is in units of the fundamental frequency $\omega = 0.167$~a.u. Blue/purple arrows indicate absorption of a pump photon $7\omega$/ $9\omega$, while red arrows indicate absorption/stimulated emission of a probe photon $\omega$, which is delayed by a time $\tau$. The RABBIT beating is measured at $\mathbf{k}_\text{SB}$. (b,c) Electron momentum distribution of the sideband (orange curve in (a)) along the $\mathbf{K}-\mathbf{M}$ ($k_\parallel$) direction for a pump-probe time delay of (b) $\tau = 0$ and (c) $\tau = 0.25$ $\omega$-cycles, and for a relative harmonic phase of $\Delta \phi = \pi$. The $\mathbf{k}_\text{SB}$ momenta is enclosed by white circles and the red-dashed line indicates the edges of the first Brillouin zone. The top left inset shows the hBN lattice with the black arrow indicating the polarization of the laser fields ($\parallel$ direction). (d) Amplitude (blue, left axis) and phase (red, right axis) of the two-photon Berry connections (see text for details). The black dotted lines indicate $\mathbf{k}_\text{SB}$. (e) Sideband population normalized to the average for four choices of harmonic phase $\Delta \phi = 0, \pi/2, \pi, 3\pi/2$ (light to dark red). The average sideband population $\bar{\rho}_{\text{SB}} = \sum_{n} \rho_{\text{SB}} (\tau_{n})/n$, being $n$ the number of time delays considered.} 
\end{center}
\end{figure*}

We first illustrate the standard RABBIT scheme for the extraction of the relative harmonic phase. This scheme involves the frequencies $(2N\pm 1)\omega$ and $\omega$, which requires three conduction bands separated by $\omega$ at a given sideband momentum $\mathbf{k}_{\text{SB}}$ (see Fig.~\ref{fig:hBN}a).

Fig.~\ref{fig:hBN}a shows the band structure of hBN along the $\mathbf{K}$-$\mathbf{M}$ direction, where we have included the two $p_z$ orbitals and the six $sp_2$ orbitals (see Appendix A for details). We choose this direction because it is inversion symmetric, and thus the dipole moments are real. This automatically eliminates the ``atomic phase" contribution to the RABBIT beating (apart from an overall $N\pi$ phase, with $N$ an integer), and allows to extract the relative phase of the harmonics in a clean way. At the crystal momentum $\mathbf{k}_\text{SB}$, where the blue and red arrows are located in Fig.~\ref{fig:hBN}a, the $sp_2$ conduction bands are separated by roughly the third harmonic of an 800nm Ti:Sapph laser, $\lambda \approx 270$~nm.
To perform the RABBIT scheme, we use the fundamental frequency $\omega = 0.167$~a.u. ($\lambda = 270$~nm), and its 7th (21st of 800nm) and 9th (27th of 800nm) harmonic. The harmonics are resonant with the lowest and highest $sp_2$ conduction bands from the lowest $sp_2$ valence band at $\mathbf{k}_\text{SB}$ (blue and purple arrows in Fig.~\ref{fig:hBN}a). The time-delayed probe, carried at $\omega$, is resonant with the intermediate $sp_2$ band (the sideband, orange curve in Fig.~\ref{fig:hBN}a) from the other two $sp_2$ conduction bands at $\mathbf{k}_\text{SB}$. Since all the valence bands are fully occupied before the pump photon arrives, the sideband is populated predominantly via two quantum paths: (i) by the absorption of a $7\omega$ photon, followed by the absorption of a $\omega$ photon, (ii) by the absorption of a $9\omega$ photon, followed by the stimulated emission of a $\omega$ photon (Fig.~\ref{fig:hBN}a). The pump fields have a strength of $F_{7\omega} = F_{9\omega} = 0.01$~V/\AA~ and the probe pulse has a strength $F_{\omega} = 0.02$~V/\AA. All fields are 15~fs long and linearly polarized along the $\mathbf{K}$-$\mathbf{M}$ direction.

In Fig.~\ref{fig:hBN}b,c we show the $\mathbf{k}$-resolved electron population of the sideband in a slice of the first Brillouin zone along the $\mathbf{K}$-$\mathbf{M}$ direction for two time delays separated by half the RABBIT period and for a relative harmonic phase of $\phi_{9\omega}-\phi_{7\omega} = \pi$. The sideband momentum $\mathbf{k}_\text{SB}$ is encircled by the white line and shows maximum electron population for panel (b) and a minimum for panel (c). Fig.~\ref{fig:hBN}e shows the oscillating population, obtained from averaging the k-resolved populations over a circle centered at $\mathbf{k}_\text{SB}$ and with 0.005 a.u. radius. Several choices of the relative harmonic phase $\Delta \phi$ are shown. As a function of the pump-probe time delay, the populations oscillate following the characteristic RABBIT frequency. Since the ``atomic" phase $\Delta \varphi=\pi$, regardless of which frequency we choose (see Fig.~\ref{fig:hBN}d), the RABBIT beating signal reconstructs directly $\Delta \phi$ with the offset of $\pi$. This should be contrasted to the atomic case, where $\Delta \varphi$ must be approximated or calculated theoretically to retrieve $\Delta \phi$, since the former varies with frequency and atomic species and is not in general a multiple of $\pi$~\cite{Paul2001}.

\subsection*{Reconstruction of the interband phase}

\begin{figure*}
\begin{center}
\includegraphics[width=\linewidth]{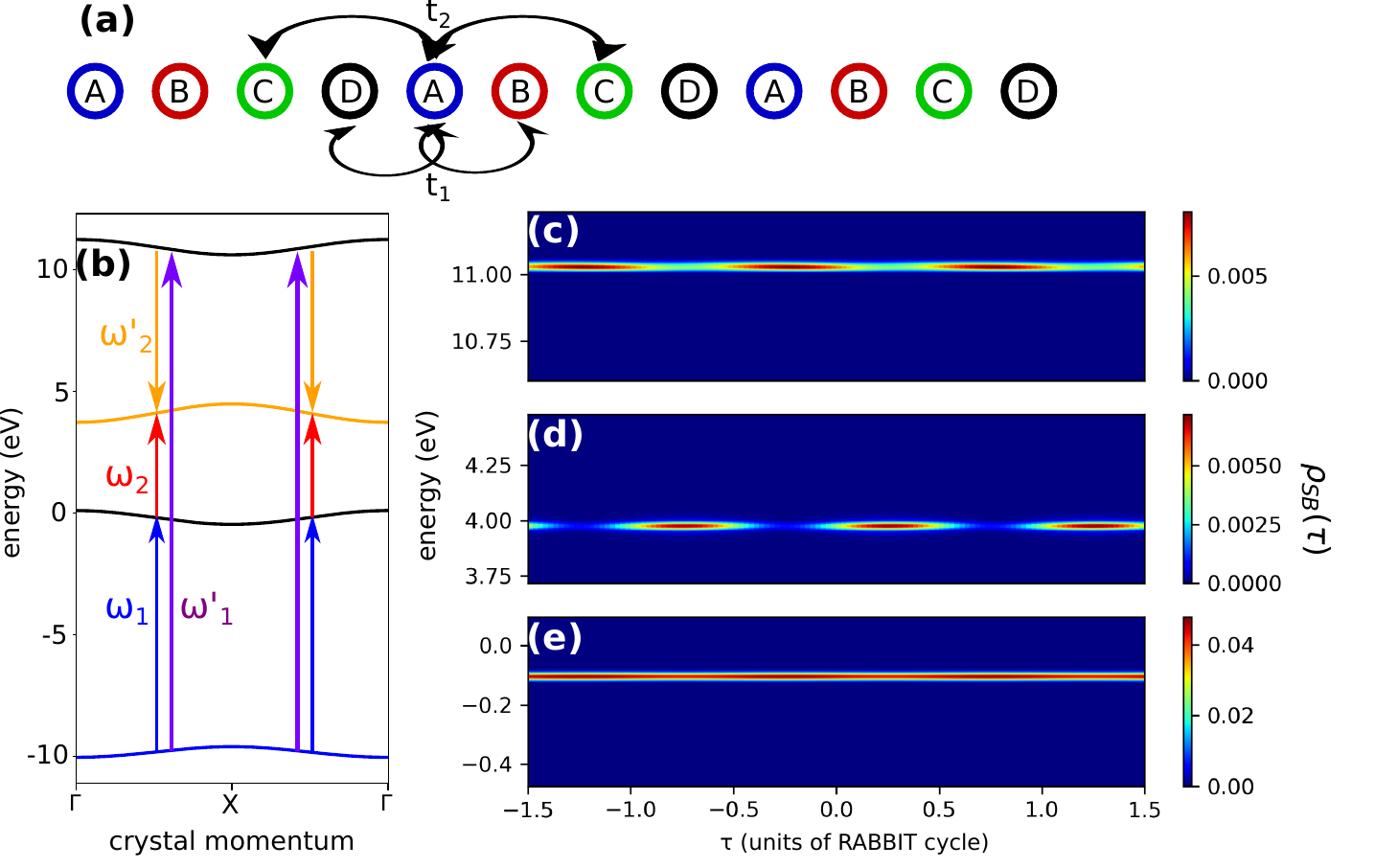}
\caption{\label{fig:model}
(a) Pictorial representation of the 1D model system (see text for details). (b) Band structure of the 1D model system with the arrows indicating the general RABBIT mechanism with four different frequencies: the pump $\omega_1$ and $\omega_1'$, and the probe $\omega_2$ and $\omega_2'$. 
(c-e) Population as a function of the pump and probe time delay at the energy of the upper conduction band (c), the sideband (d, RABBIT beating), and the lower conduction band (e).}
\end{center}
\end{figure*}

Next we illustrate the case where the relative harmonic phase is known in advance and one is interested in extracting the phase information of interband transitions~\cite{Dahlstrom2012}. The matrix element between the Bloch state of band $m$ and that of band $n$ at the crystal momentum $\mathbf{k}$ can be written in terms of the position operator,
\begin{equation}
    \hat{\mathbf{r}} = i \partial_{\mathbf{k}} \delta_{nm} + \mathbf{A}_{nm\mathbf{k}}.
\end{equation}
The first term in the RHS can be neglected for weak fields, since it accounts for the ``streaking" of electrons by the vector potential, which is weak in the RABBIT method. The second term, $\mathbf{A}_{nm\mathbf{k}} = i\,\langle u_{n\mathbf{k}} | \nabla_{\mathbf{k}} u_{m\mathbf{k}} \rangle$ is the Berry connection, with $u_{n\mathbf{k}}$ the periodic part of the Bloch state of band $n$. Then, the two-photon matrix element between an initial state in band $m$ and a final state in sideband $n$ at the crystal momentum $\mathbf{k}$ is the two-photon Berry connection (TPBC),
\begin{equation}
    {A}^{(2)}_{nm\mathbf{k},\parallel} (\omega) = \sum_{j} {A}_{nj\mathbf{k},\parallel} {A}_{jm\mathbf{k},\parallel} / \left[\varepsilon_m(\mathbf{k}) + \omega - \varepsilon_j (\mathbf{k})\right],
\end{equation}
where $\parallel$ indicates the component parallel to the laser polarization (we assume all fields linearly polarized along the same direction). The ``atomic" phase recorded by RABBIT in interband transitions is the relative phase between the TPBC,
\begin{equation}\label{eq:atomic_phase_solids}
\begin{split}
\Delta \varphi_{nm\mathbf{k}}^{\text{(inter)}} &= \arg \left\{{A}^{(2)}_{nm\mathbf{k},\parallel}\left((2N+1)\omega\right) + {A}^{(2)}_{nm\mathbf{k},\parallel} \left(-\omega\right) \right\} \\
& - \arg \left\{{A}^{(2)}_{nm\mathbf{k},\parallel}\left((2N-1)\omega\right) + {A}^{(2)}_{nm\mathbf{k},\parallel} \left(\omega\right) \right\}.
\end{split}
\end{equation}
The analogy to the atomic case shows that it is possible to extract information on interband dynamics in solids in the same way as photo-ionization dynamics in atoms. We will refer to the ``atomic phase" in solids (Eq.~\ref{eq:atomic_phase_solids}) as the interband phase. Retrieving this phase is of interest in materials with broken inversion symmetry and along non-inversion-symmetric directions, where it will not simply be a multiple of $\pi$ as we have seen before.

To illustrate this, we consider a one-dimensional chain made up of four different atoms (A,B,C,D) placed consecutively, each with one atomic orbital, such that the system has broken inversion symmetry (Fig.~\ref{fig:model}a). We choose the distances between atoms to be 2, 4, 6 and 8 a.u. for A-B, A-C, A-D and A-A, respectively. We consider first and second neighbour hoppings ($|t_1| = 3$~eV, $|t_2| = 1$~eV), and on-site energies $\Delta = -8$, $0$, $5$ and $8$ eV for atoms A,B,C and D, respectively. The band structure is shown in Fig.~\ref{fig:model}b. Since we are interested in exploring the phase of the material throughout the Brillouin zone, in this case it is necessary to assume that we have access to four different phase-locked tunable frequencies. While this is technically challenging, it is in principle possible to achieve through high harmonic generation of both a signal and an idler wavelength.

The scheme thus assumes the general case in which for each momentum $\mathbf{k}_{SB}$, we have two pump frequencies $\omega_1$ and $\omega_1'$ and two probe frequencies $\omega_2 $ and $\omega_2'$ such that at the sideband energy $\Omega \equiv \omega_1'-\omega_2' = \omega_1 + \omega_2$ (see Fig.~\ref{fig:model}b and Methods). Fig.~\ref{fig:model}c-e show the populations of the three conduction bands as a function of the time delay between pump and probe frequencies, for the choice of frequencies depicted in panel b, i.e., resonant at $\mathbf{k}_{SB}$. The signal is measured for $\mathbf{k}>0$, i.e., along the $\Gamma$-X direction. A momentum-resolved measurement is required because, due to the absence of inversion symmetry $\Delta \varphi_{\mathbf{k}_{SB}}^{\text{(inter)}} = -\Delta \varphi_{-\mathbf{k}_{SB}}^{\text{(inter)}}$. The energy range covered by the panels c-e corresponds to the bandwidth of each of the conduction bands. The bands are populated at the energy corresponding to the crystal momentum $\mathbf{k}_{SB}$. The sideband (panel d) shows the characteristic RABBIT beating, while a third-order process modulation is also visible in the upper band (panel c). Provided the relative harmonic phase $\Delta \phi$ is known, the phase of the sideband beating allows to extract the interband phase at $\mathbf{k}_{SB}$, $\Delta\varphi_{\mathbf{k}_{SB}}^{\text{(inter)}}$.

\begin{figure*}
\begin{center}
\includegraphics[width=\linewidth]{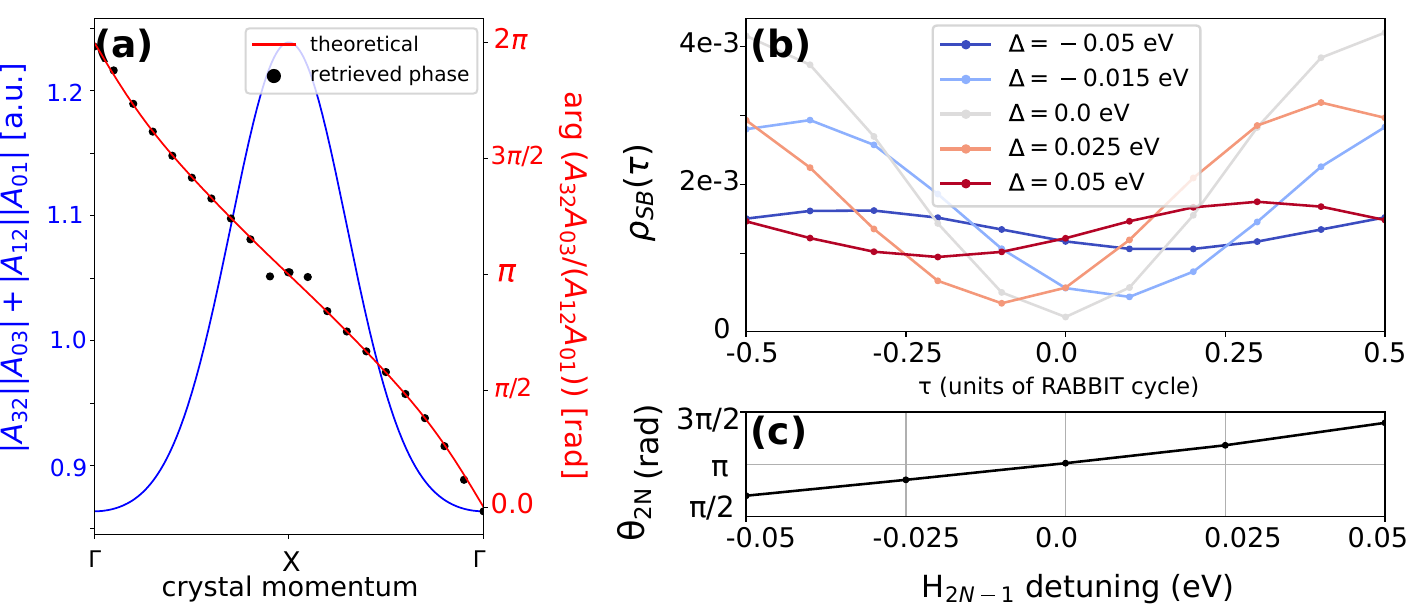}
\caption{\label{fig:reconstruction} 
(a) Amplitude of the two-photon Berry connection (blue curve, left axis) and interband phase (red curve, right axis) of the model in Fig.~\ref{fig:model}. The black points indicate the phase reconstructed with the RABBIT method at each crystal momenta. (b) Sideband population beating at $\mathbf{k}_{SB} =$ X for various frequency detunings between $\omega_1$ and the energy of the first conduction band at X (see Fig.~\ref{fig:model}b): -0.05eV in dark blue, -0.025eV in light blue, on resonance in grey, 0.025eV in light red and 0.05eV in dark red. (g) RABBIT phase of the beatings in panel (f) as a function of the detuning, extracted from fitting to Eq.~\ref{eq:beating}.}
\end{center}
\end{figure*}

Fig.~\ref{fig:reconstruction}a demonstrates the interband phase reconstruction along the full Brillouin zone. To obtain this, we performed a RABBIT measurement at each crystal momenta by tuning the four frequencies $\omega_1$, $\omega_1'$, $\omega_2$ and $\omega_2'$ in Fig.~\ref{fig:model}b. For each measurement, we extracted the phase of the oscillation (Fig.~\ref{fig:model}d). These phases are shown as black points in Fig.~\ref{fig:reconstruction}a, and  faithfully reproduce the theoretical values computed through Eq.~\ref{eq:atomic_phase_solids} (red curve).

\subsection*{Influence of harmonic-band detuning in the RABBIT phase}

One of the most striking examples of the link between the RABBIT beating phase and the phase of the two-photon matrix elements in atoms and molecules is that in which one of the harmonics traverses a bound state~\cite{Swoboda2010, Caillat2011, Jimenez2014, Kotur2016, Gruson2016, Cirelli2018}. In this case, the intermediate step in the two-photon process is dominated by a single resonant state with energy $E_{\text{res}}$, so that the two-photon matrix element $\mathcal{M}_{nm} (\omega) = \mathcal{O}_{nj}\mathcal{O}_{jm} / (E_m + \omega - E_\text{res})$ displays a phase shift of $\pi$ as a function of the detuning of the frequency $\omega$ from the resonance~\cite{Swoboda2010}. 

To test this process in solids, we consider again the 1D chain system in Fig.~\ref{fig:model}. We change the frequency of $\omega_1$, while keeping all other field parameters the same; in particular, the frequencies $\omega_1'$, $\omega_2$ and $\omega_2'$ remain unchanged. Such conditions are chosen to illustrate the effect. However, we note that the different curvature that each band has as a function of $\mathbf{k}$ will likely give rise to a similar scenario when the generating frequency $\omega$ is changed, as in previous works~\cite{Kotur2016}.

Fig.~\ref{fig:reconstruction}b shows the RABBIT beating and phase for various detunings between $\omega_1$ and the first conduction band at $\mathbf{k}_{SB}=$ X. Far from resonance (dark red and dark blue), the RABBIT beating is small, but one can clearly distinguish a $\pi$ shift in the oscillation between negatively detuned (dark blue) and positively detuned (dark red) frequencies. To visualize it more clearly, we fit the oscillations to Eq.~\ref{eq:beating} and extract the RABBIT phase. Fig.~\ref{fig:reconstruction}c shows the result, where one can observe the $\pi$ shift as $\omega_1$ crosses the resonance, in agreement with Eq.~\ref{eq:atomic_phase_solids}. Note that higher/lower detunings than those shown will not generate a RABBIT oscillation.

\subsection{Extraction of electron-hole dephasing times by non-overlapping pump and probe pulses}

So far, we have demonstrated how the RABBIT technique can be applied to solids, allowing to extract the same dynamical observables as in atoms or molecules. There is, however, a fundamental difference with respect to the traditional RABBIT implementation in atoms. In atoms, if the pump and the probe pulses do not temporally overlap, radiative transitions between unstructured continuum states, i.e., absent of resonances, are forbidden. In solids, the interband transitions are resonant and thus always allowed: there will be a RABBIT beating for non-overlapping pulses as long as there is coherence between the bands.

When dephasing can be well approximated by a constant exponential decay of the coherence, the RABBIT beating amplitude will simply follow the function
\begin{equation}\label{eq:fit}
f(t) = A e^{-t/T_2} \cos \left[2\omega t + \theta\right] + B,
\end{equation}
where $A$, $B$, $\theta$ and $T_2$ are fitting parameters (many of which can be strongly bounded). For more complex decoherence mechanisms, the fitting function may need to include more parameters, but the general approach still remains valid.

To study the dependence of the RABBIT beating on electron decoherence, we introduce a dephasing time in the 1D chain model (Fig.~\ref{fig:model}). As it is commonly done in other works~\cite{Vampa2014}, we do so with a phenomenological parameter $T_2$ that exponentially suppresses the non-diagonal elements of the density matrix (see Appendix B for further details). Fig.~\ref{fig:decoherence} shows the RABBIT beating in the 1D chain model with $T_2=12$~fs for a wide range of pump-probe time delays. For $\tau=0$ (nor shown), the pulses are perfectly overlapping, and they start to fully separate for $\tau>22$~fs. The RABBIT signal is computed at $\mathbf{k}_{SB} =$ X, but other momenta show the same results. To reduce computational cost, the duration of the pulses in this case was limited to 8~fs full width at half maximum. The field strengths were $F_{\omega_1'} = F_{\omega_1} = 0.01$~V/\AA, $F_{\omega_2} = F_{\omega_2'} = 0.02$~V/\AA. The beating amplitude in the non-overlap region follows an exponential decay as a consequence of electronic decoherence. Fitting the beating to Eq.~\ref{eq:fit} yields a value of the dephasing parameter of $T_2^{\text{(fit)}} = 12.09$~fs, in perfect agreement with the numerical input value of $T_2 = 12$~fs. For these simulations we have neglected the effect of population relaxation (which will lead to a similar decay of the RABBIT signal) since its timescale is generally much larger than that of dephasing.

\begin{figure}
\begin{center}
\includegraphics[width=\linewidth]{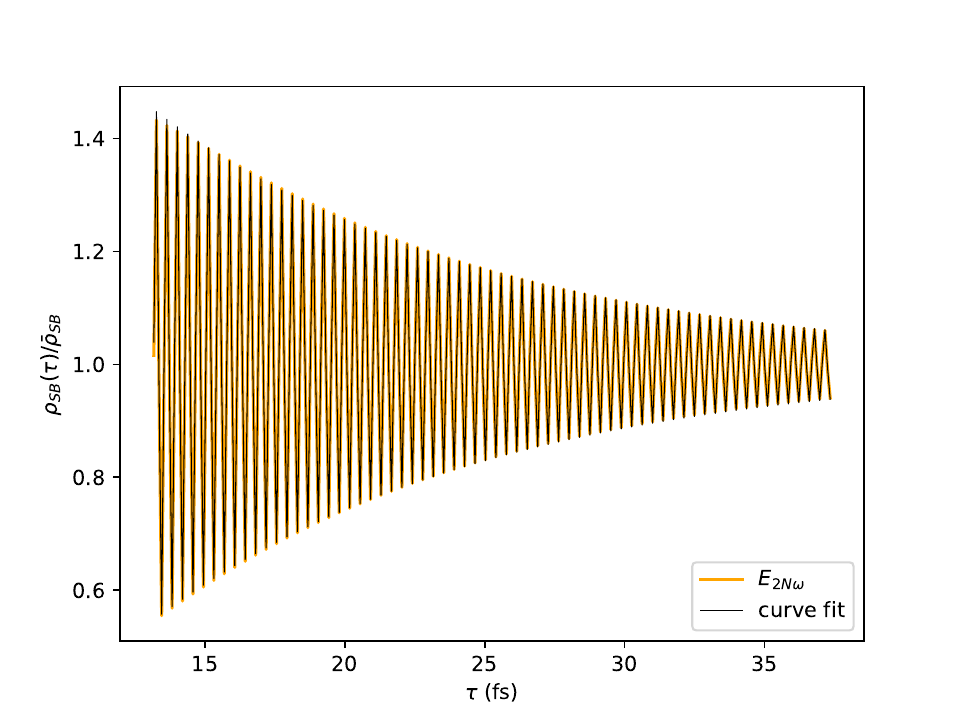}
\caption{\label{fig:decoherence} (a) In orange, the RABBIT signal at $\mathbf{k}_{SB} =$ X for different pump-probe time delays $\tau$ (non-overlapping starts at $\tau > 22$~fs). The dephasing time is $T_2=12$~fs. In black, the fitting to Eq.~\ref{eq:fit}, which yields $T_2^{\text{(fit)}} = 12.09$~fs.}
\end{center}
\end{figure}

\section*{Conclusion}

To summarize, we have explored application of the RABBIT technique to interband transitions in solids by measuring the momentum-resolved sideband population, which is possible through ARPES. Similar to its application in atoms, this may open the way to obtain relative amplitudes and phases of two-photon dipole couplings, which are hard to obtain even through numerical methods in most condensed matter systems~\cite{Marzari2012,Silva2019-maximally}, thus giving a window into the dynamics of interband excitations. Using a toy model, we have shown that the relative phase of the harmonics can be retrieved directly from the sideband population beating in inversion-symmetric systems (or along inversion-symmetric directions), without the influence of the interband (``atomic") phase, in contrast to atoms.  Finally, we have shown that the RABBIT signal in solids is extremely sensitive to decoherence mechanisms, providing a simple interferometric way to extract dephasing times.

\section*{Acknowledgements}
We thank Misha Ivanov for illuminating discussions and Luca Argenti for sharing his enlightening view of the RABBIT technique. This work was funded by the European Union’s Horizon 2020 research and innovation programme under the Marie Skłodowska-Curie grant agreement no. 101028938. R.E.F.S. also acknowledges support from the fellowship Grant No. LCF/BQ/PR21/11840008 from “LaCaixa” Foundation (ID 100010434).

\section*{Appendix A: hBN field-free DFT calculation}

To obtain the field-free band structure and dipole couplings (Berry connections) of hBN we first perform a DFT calculation with Quantum Espresso~\cite{Gianozzi2009}. We use a HSE functional with a 10x10x4 Monkhorst-Pack grid. This gives a minimum band gap of $\sim 6$~eV, in line with previous works. To achieve a fixed phase relation of the dipole couplings at different $k$, we transform our basis to the maximally-localized Wannier basis with the Wannier90 suite~\cite{Marzari2012}. For this, we project onto the $p_z$ and $sp_2$ orbitals of boron and nitride. In this way, we obtain a tight-binding representation of field-free hBN, which is then propagated using the code~\cite{Silva2019-maximally}, briefly described below.

\section*{Appendix B: Propagation in presence of the field} We solve the density matrix equation in the dipole approximation and in the length gauge,
\begin{equation}\label{eq:density_matrix}
\partial_t \rho_{nm} (\mathbf{k}, t) = -i \left[ \hat{H}(\mathbf{k},t), \hat{\rho}(\mathbf{k},t)\right]_{nm} - \frac{(1-\delta_{nm})\rho_{nm}(\mathbf{k},t)}{T_2}.
\end{equation}
The Hamiltonian of our system is $H(\mathbf{k},t) = H_0(\mathbf{k}) + |e| \mathbf{E}(t) \cdot \mathbf{r}$, where $H_0$ is the periodic field-free tight-binding Hamiltonian constructed as indicated above and $\mathbf{E}(t)$ is the time-dependent field. The representation of the position operator is that given by Blount~\cite{blount1962}, $\mathbf{\hat{r}} = i \partial_{\mathbf{k}} +\mathbf{\hat{A}}(\mathbf{k})$, where $\mathbf{\hat{A}}(\mathbf{k})$ is the Berry connection. Dephasing is introduced in a phenomenological way via the constant dephasing time parameter $T_2$, which exponentially suppresses the coherences between the bands. The initial state is a mixed state, with no coherence between the bands, where the valence band is fully occupied and the conduction bands are empty. The final populations are obtained from the diagonal elements of the density matrix at a time when the pulse is over. For the 1D chain model, we used a grid of $N_{k} = 600$ points and a step size of $dt = 0.1$~a.u.. For monolayer hBN, we used a grid of $N_{k_x} = N_{k_y} = 400$ points and a $dt=0.1$~a.u. Further details of the code can be found in~\cite{Silva2019-maximally}.

\section*{Appendix C: RABBIT expression}

The RABBIT protocol is an interference of four time-ordered two-photon paths that reach the same final energy $\Omega$. Since we consider transitions from a low-lying state to a high-lying state, e.g., ground electronic state to continuum or deep valence band to conduction band, we assume that the paths are formed by the absorption of a high frequency pump photon and the absorption or stimulated emission of a probe photon. Let us denote the pump photons by $\omega_1$ and $\omega_1'$ and the probe photons by $\omega_2$ and $\omega_2'$. The paths interfere at the energy $\Omega$. To reproduce the usual RABBIT implementation, where there are two paths contributing ``from above" and two ``from below" (Fig.~\ref{fig:model}b), we take $\Omega = \omega_1+\omega_2 = \omega_1'-\omega_2'$, where $+$ corresponds to photon absorption and $-$ to stimulated photon emission; any other combination of frequencies will not contribute to this energy, e.g., $\omega_1+\omega_2' \neq \Omega$. Therefore, the paths containing $\omega_1 < \Omega$ contribute ``from below" and are associated to $\omega_2$ probe photon absorption while the paths containing $\omega_1' > \Omega$ contribute ``from above" and are associated to $\omega_2'$ probe photon emission. We write the four paths as: (i) $+\omega_1  + \omega_2$, (ii)  $+\omega_1' - \omega_2'$, (iii) $+\omega_2 + \omega_1$, (iv) $-\omega_2' + \omega_1'$. The first frequency is absorbed/emitted first. 

Let us consider path (i), where we first absorb the pump photon $\omega_1$, and then absorb the probe photon $\omega_2$. We assume monochromatic pulses, so that for the pump field we have
\begin{equation}
F_{\omega_1}(t) = \frac{F_{0,\omega_1}}{2}\left( e^{i\phi_{\omega_1}} e^{i\omega_1 t} + e^{-i\phi_{\omega_1}} e^{-i\omega_1 t} \right).
\end{equation}
The probe field is delayed by a time $\tau$, so that
\begin{equation}
F_{\omega_2}(t) = \frac{F_{0,\omega_2}}{2}\left( e^{i(\phi_{\omega_2}+\omega_2\tau)} e^{i\omega_2 t} + e^{-i(\phi_{\omega_2}+\omega_2 \tau)} e^{-i\omega_2 t} \right).
\end{equation}
The second order amplitude in the Dyson expansion is
\begin{equation}
\begin{split}
\mathcal{A}_{fi}^{(2)} =& - \sum_j \mathcal{O}_{fj} \mathcal{O}_{ji} \int_{t_0}^{t} dt_1\,e^{iE_{fj}t_1} F(t_1) \times \\
&\int_{t_0}^{t_1} dt_2\,e^{iE_{ji} t_2}\,F(t_2),
\end{split}
\end{equation}
where the $\mathcal{O}_{ba}$ are the transition matrix elements, e.g., in length gauge $\mathcal{O}_{ba} = \langle b | \mathbf{r} | a \rangle$. For monochromatic pulses, we can compute the amplitude taking $t_0 \to -\infty$ and $t\to +\infty$.  Then,
\begin{equation}\label{eq:secorder}
\begin{split}
\mathcal{A}_{fi,\text{(i)}}^{(2)} =& - \frac{F_{0,\omega_2}e^{-i(\phi_{\omega_2}+\omega_2 \tau)}}{2} \frac{F_{0,\omega_1}e^{-i\phi_{\omega_1}}}{2}\sum_j \mathcal{O}_{fj} \mathcal{O}_{ji} \times \\
&\int_{-\infty}^{\infty} dt_1\,e^{i (E_{fj}-\omega_2) t_1} \int_{-\infty}^{t_1} dt_2\,e^{i(E_{ji}-\omega_1) t_2},
\end{split}
\end{equation}
where subscript $\text{(i)}$ indicates which path we consider. The second integral in Eq.~\ref{eq:secorder} can be solved by introducing a decaying exponential in the limit of $t\to \infty$,
\begin{equation}\label{eq:distribution}
\int_{-\infty}^{t} d\tau e^{i\omega \tau} \to \lim_{\nu\to 0^+} \int_{-\infty}^{t} dt_1 e^{i(\omega-i\nu)t_1} = i \frac{e^{i\omega t}}{-\omega + i0^+},
\end{equation}
where $(x+i0^+)^{-1}$ is a distribution. Using Eq.~\ref{eq:distribution} in the second integral of Eq.~\ref{eq:secorder},
\begin{equation}
\begin{split}
\mathcal{A}_{fi,\text{(i)}}^{(2)} =& - i \frac{F_{0,\omega_2}e^{-i(\phi_{\omega_2}+\omega_2 \tau)}}{2} \frac{F_{0,\omega_1}e^{-i\phi_{\omega_1}}}{2} \times \\
&\sum_j \frac{\mathcal{O}_{fj} \mathcal{O}_{ji}}{E_i + \omega_1 - E_j + i0^+} \int_{-\infty}^{\infty} dt_1\,e^{i (E_{fi}-\omega_1 - \omega_2) t_1}.
\end{split}
\end{equation}
We identify the integral with the delta distribution, i.e., $\int_{-\infty}^{\infty} dt\,e^{i(\omega-\omega_0) t} = 2\pi\delta(\omega - \omega_0)$, so that
\begin{equation}
\begin{split}
\mathcal{A}_{fi,\text{(i)}}^{(2)} =& -  \frac{i \pi\, F_{0,\omega_2}F_{0,\omega_1}e^{-i(\phi_{\omega_2}+\omega_2 \tau)}e^{-i\phi_{\omega_1}}}{2} \times \\
&\sum_j \frac{\mathcal{O}_{fj} \mathcal{O}_{ji}}{E_i + \omega_1 - E_j + i0^+}\,\delta(E_{fi}-\omega_1-\omega_2).
\end{split}
\end{equation}
For compactness, let us write the two-photon transition matrix element as
\begin{equation}
\mathcal{M}_{fi} (\omega_1) = \sum_j \frac{\mathcal{O}_{fj} \mathcal{O}_{ji}}{E_i + \omega_1 - E_j + i0^+}.
\end{equation}
Then, the transition amplitude reads,
\begin{equation}\label{eq:transition_amplitude_rabbit}
\mathcal{A}_{fi,\text{(i)}}^{(2)} = - \frac{i\pi\,F_{0,\omega_2}F_{0,\omega_1}e^{-i(\phi_{\omega_2}+\phi_{\omega_1}+\omega_2 \tau)}}{2}\,\mathcal{M}_{fi}(\omega_1)\,\delta(E_{fi}-\Omega),
\end{equation}
where $\Omega = \omega_1 + \omega_2 = \omega_1' - \omega_2'$ is the sideband energy. The other three paths contributing to the RABBIT signal follow,
\begin{equation}
\mathcal{A}_{fi,\text{(ii)}}^{(2)} = - \frac{i\pi\,F_{0,\omega'_2}F_{0,\omega_1'}e^{i(\phi_{\omega_2'}-\phi_{\omega_1'}+\omega_2' \tau)}}{2}\,\mathcal{M}_{fi}(\omega'_1)\,\delta(E_{fi}-\Omega),
\end{equation}
\begin{equation}
\mathcal{A}_{fi,\text{(iii)}}^{(2)} = - \frac{i\pi\,F_{0,\omega_2}F_{0,\omega_1}e^{-i(\phi_{\omega_2}+\phi_{\omega_1}+\omega_2 \tau)}}{2}\,\mathcal{M}_{fi}(\omega_2)\,\delta(E_{fi}-\Omega),
\end{equation}
\begin{equation}
\mathcal{A}_{fi,\text{(iv)}}^{(2)} = -  \frac{i\pi\,F_{0,\omega'_2}F_{0,\omega_1'}e^{i(\phi_{\omega_2'}-\phi_{\omega_1'}+\omega_2' \tau)}}{2}\,\mathcal{M}_{fi}(-\omega'_2)\,\delta(E_{fi}-\Omega).
\end{equation}
We group paths (i) and (iii) on the one hand, and paths (ii) and (iv) on the other, which correspond, respectively, to absorption of an $\omega_2$ photon and stimulated emission of an $\omega'_2$ probe photon,
\begin{equation}\label{eq:secorder_one}
\begin{split}
\mathcal{A}_{fi,\text{em}}^{(2)} =& -  \frac{i\pi\,F_{0,\omega_2}F_{0,\omega_1}e^{-i(\phi_{\omega_2}+\phi_{\omega_1}+\omega_2 \tau)}}{2} \times \\
&\left[\mathcal{M}_{fi}(\omega_1) + \mathcal{M}_{fi}(\omega_2) \right]\,\delta(E_{fi}-\Omega).
\end{split}
\end{equation}
\begin{equation}\label{eq:secorder_two}
\begin{split}
\mathcal{A}_{fi,\text{abs}}^{(2)} =& - \frac{i\pi\,F_{0,\omega_2'}F_{0,\omega_1'}e^{i(\phi_{\omega_2'}-\phi_{\omega_1'}+\omega'_2 \tau)}}{2} \times\\
&\left[\mathcal{M}_{fi}(\omega'_1) + \mathcal{M}_{fi}(-\omega'_2) \right]\,\delta(E_{fi}-\Omega).
\end{split}
\end{equation}
The total intensity at the sideband energy is the coherent sum of Eq.~\ref{eq:secorder_one} and Eq.~\ref{eq:secorder_two},
\begin{equation}\label{eq:rabbit_four_freq}
\begin{split}
I_{\text{SB}} =& | \mathcal{A}_{fi,\text{abs}}^{(2)} + \mathcal{A}_{fi,\text{em}}^{(2)} |^2 \propto\\
& \cos \left[ \Delta\phi - (\phi_{\omega_2} + \phi_{\omega'_2}) + \Delta \varphi - (\omega_2 + \omega_2') \tau\right],
\end{split}
\end{equation}
where we have defined the relative harmonic phase, $\Delta {\phi} \equiv \phi_{\omega'_1} - \phi_{\omega_1}$, and the interband (``atomic") phase
\begin{equation}\label{eq:phases_four_freq}
\begin{split}
&\Delta \varphi \equiv \varphi_{\text{abs}} - \varphi_{\text{em}}, \\
&\varphi_{\text{abs}} = \arg [\mathcal{M}_{fi}(\omega_1) + \mathcal{M}_{fi}(\omega_2)], \\
&\varphi_{\text{em}} = \arg [\mathcal{M}_{fi}(\omega'_1) + \mathcal{M}_{fi}(-\omega'_2)].
\end{split}
\end{equation}
For the case in which the probe frequencies are the same $\omega_2 = \omega_2' = \omega$, and the pump frequencies are odd harmonics of the probe, $\omega_1 = (2N-1)\omega$ and $\omega'_1 = (2N+1)\omega$, Eq.~\ref{eq:rabbit_four_freq} and Eq.~\ref{eq:phases_four_freq} reduce to Eq.~\ref{eq:beating} and Eq.~\ref{eq:atomic_phase}.



\bibliography{biblio}

\end{document}